\documentclass[aps,pra,twocolumn,showpacs,superscriptaddress]{revtex4} \usepackage{graphicx,amssymb,amsfonts,amsmath}

\usepackage{ulem}
\usepackage{framed}

\newcommand{\br}{\mathbf{r}}
\newcommand{\bk}{\mathbf{k}}
\newcommand{\etal}{~\text{et. al.}}
\newcommand{\CO}{(Color online)\;}

\newcommand{\bp}{\begin{pmatrix}}
\newcommand{\ep}{\end{pmatrix}}

\newcommand{\be}{\begin{eqnarray}}
\newcommand{\ee}{\end{eqnarray}}
\newcommand{\vk}{{\bf v_k}}
\newcommand{\vkt}{{\bf v}^T_\bk}
\newcommand{\ek}{\epsilon_{\bf k}}
\newcommand{\fk}{f_{\bf k}}
\newcommand{\bphi}{\boldsymbol{\phi}}
\newcommand{\n}{\nonumber \\}


\usepackage[utf8]{inputenc} 
\usepackage{graphicx}
\usepackage{bbold}

\begin{document}

\title{
Damping of Bloch oscillations: variational solutions \\of the Boltzmann equation beyond linear response
}
\author{Stephan Mandt}
\email{smandt@princeton.edu}
\affiliation{Princeton Center for Complex Materials, Princeton University, New Jersey 08544, USA}
\date{August 31, 2014}

\begin{abstract}
Variational solutions of the Boltzmann equation usually rely on the concept of linear response. 
We extend the variational approach for tight-binding models at high entropies to a regime far beyond linear response. We analyze both weakly 
interacting fermions and incoherent bosons on a lattice. We consider a case where the particles are driven by a constant force,
leading to the well-known Bloch oscillations, and we consider interactions that are weak enough not to overdamp these oscillations.
This regime is computationally demanding and relevant for ultracold atoms in optical lattices.
We derive a simple theory in terms of coupled dynamic equations for the particle density, energy density, current and heat current, allowing 
for analytic solutions. As an application, we identify damping coefficients for 
Bloch oscillations in the Hubbard model at weak interactions and compute them for a one-dimensional toy model. We also approximately solve the long-time dynamics of a weakly interacting,
strongly Bloch-oscillating cloud of fermionic particles in a tilted lattice, leading to a subdiffusive scaling exponent.
\end{abstract}

\maketitle

\section{Introduction}

Boltzmann equations are established tools to study the nonequilibrium dynamics of electron gases in materials~\cite{ziman}
and of atomic quasi-particles in traps~\cite{griffin}. They describe the time evolution of a phase-space probability distribution of particles subject to external forces and collisions,
e.g., 
due to disorder, phonons or interactions.   
In most applications in condensed-matter or material physics, the electrical field that drives the electronic system out of equilibrium
is weak. In this case, 
linearizing the Boltzmann equation around a local equilibrium solution delivers transport and linear response quantities such as electronic or thermal conductivities
and thermoelectric coefficients; for a general overview see Ref.~\cite{ziman}.
In spatially inhomogeneous situations, the linearized Boltzmann equation can be used to calculate diffusion constants, allowing us to model the  flow of mass and
energy in spatially inhomogeneous setups; an application with ultracold atoms in optical lattices is given in Ref.~\cite{expansion}. 

Analytical solutions of the Boltzmann equation beyond the linear-response regime are usually infeasible, and one has to resort to simulation.
These simulations can be computationally demanding:  a high-dimensional collision integral has to be evaluated numerically for each point in phase space and time.
To circumvent expensive dynamical simulations, we introduce a novel variational approach for lattice systems. As we explain in this paper, we linearize 
the Boltzmann equation around a constant solution. This linearization is a good approximation for distribution functions at 
high entropies. 
Our approach reduces the Boltzmann equation to a coupled set of differential equations, describing the dynamics of the most relevant modes such as particle and energy densities 
and particle and heat currents. 
It is \textit{variational}, as it coincides with the conventional variational approach of solving the Boltzmann equation at high entropies in the linear response limit, as we show. 
However, the explicit time-dependence of the current and heat current allows us to model physics far beyond the linear response regime.  
We use it to derive novel analytical results with relevance to ultracold atoms in optical lattices.

\begin{figure}[ht]
\begin{center}
\includegraphics[width=\linewidth]{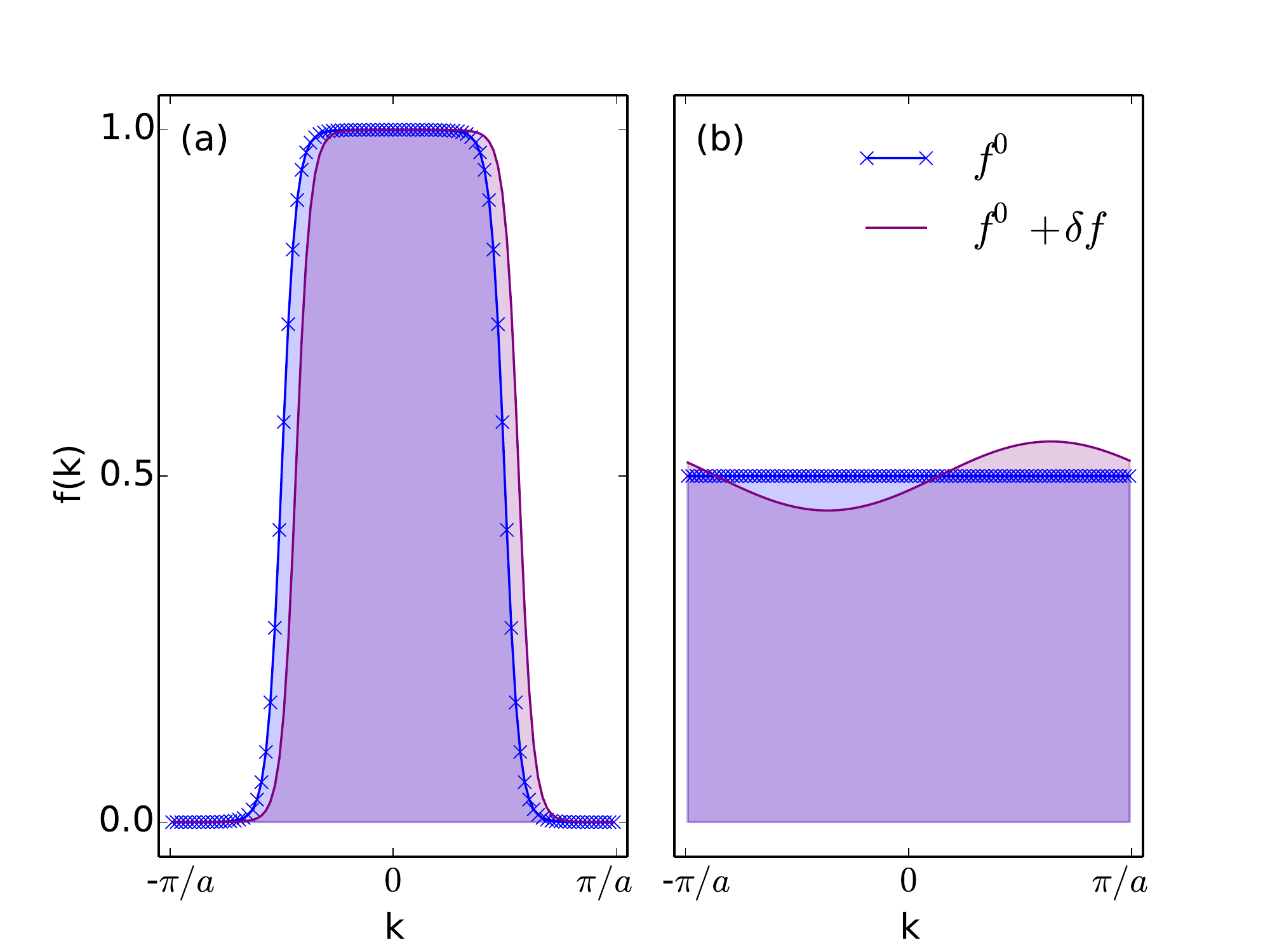}
 \caption{ 
\CO Sketch of typical momentum distribution functions $f(k)$ in equilibrium [$f^0$, blue line with crosses] and out of equilibrium [$f^0+\delta f$, solid purple line].
(a) A typical linear response setup at low temperature, or low entropy. The nonequilibrium distribution is slightly displaced to positive momenta
, e.g., due to an external force. (b) The regime of interest in this paper (high entropies and large fields). Here, the displacement to positive momenta is large.
Nevertheless, the difference $\delta f$ between the equilibrium distribution $f^0$ and the nonequilibrium distribution still remains small.
In both situations, linearization of the Boltzmann equation is possible. 
}
\label{fig:linearize}
\end{center}
\end{figure}

Using linearization techniques to model the nonlinear response regime might appear paradoxical.
Note, however,  that the term linear response usually describes
a linear relation between current and driving force. 
At low temperatures, linearization of the Boltzmann equation requires a small drive, as the difference $\delta f$ between the equilibrium distribution $f^0$ and the 
nonequilibrium distribution $f^0 + \delta f$ needs to be small. Our linearization around maximum entropy still requires $\delta f$
to be small, but it does not need to be linearly related to the driving force when the system is close enough to maximum entropy. This situation is sketched in Fig.~\ref{fig:linearize}. 
We can therefore model the nonlinear response regime while simultaneously linearizing the Boltzmann equation, which is one of the central ideas in this paper.

We use our method to model the damping of Bloch oscillations, where the current oscillates as a response to a constant drive. 
We also model an interacting atomic cloud in a tilted optical lattice and derive a stroboscopic diffusion equation
that describes its long-time dynamics. 
This regime is particularly relevant for ultracold fermionic atoms in optical lattices that are currently
only realizable at high temperatures relative to the bandwidth in an optical lattice~\cite{blochReview}, justifying the high-entropy approximation.


This paper is organized as follows. In Sec.~\ref{sec:boltzmann} we introduce our variational approach, leading to a set of coupled differential equations for 
the currents and densities. From these, we derive hydrodynamic equations in Sec.~\ref{sec:hydrodynamics} in the limit of strong damping, showing agreement with the 
conventional variational approach. We specialize our equations to the case of interacting fermions and incoherent bosons in Sec.~\ref{sec:scattering}, where we derive expressions
for the relevant scattering matrix elements. Readers primarily interested in physical results rather than the formalism may proceed to Sec.~\ref{sec:blochoscillations},
where we derive an analytic theory for the damping of Bloch oscillations at weak to intermediate interactions, including damping rates and interaction-induced frequency shifts. We also compare our analytical theory against 
a numerical simulation of the Boltzmann equation. Finally,
in Sec.~\ref{sec:stroboscopic}, we study the spatially inhomogeneous problem of an interacting, Bloch oscillating cloud in an optical lattice.
Here, we derive an approximate stroboscopic diffusion equation, which we solve asymptotically using a scaling ansatz.

\section{The Boltzmann equation at high entropies}
\label{sec:boltzmann}
We start from the Boltzmann equation. 
It describes the time evolution of a phase-space distribution function $f({\bf r},{\bf k},t)$, involving momentum ${\bf k}$,
 position ${\bf r}$, and time $t$. The left-hand side of the Boltzmann equation describes the noninteracting semiclassical 
motion and involves a force ${\bf F}$ and a group velocity ${\vk}$. The right-hand side takes collisions into account
and consists of a collision functional $I^c_{\bk}[f]$, which we leave unspecified for the moment. 
The Boltzmann equation is  given by
\begin{equation}
    \partial_t f + \mathbf{v}_\bk \cdot \nabla_\br f + \mathbf{F} \cdot \nabla_\bk f = -I^{c}_\bk[f] \;
\label{eq:Boltzmann}
\end{equation}
Throughout this text, we consider simple cubic lattices with nearest-neighbor hopping, where we set the lattice constant $a$ to $1$. This results in the following periodic dispersion:
\be
\ek = -2J\sum_{i=1}^d \cos(k_i).
\ee
Above, $d$ is the dimension of the momentum space.
The periodic energy-momentum relation is crucial to our approach; it emerges from the quantum-mechanical solution of lattice particles 
and plays an important role for ultracold atoms in optical lattices.
The corresponding group velocity $\vk = \nabla_{\bf k}\ek$ is also a nonlinear, periodic function of the quasi-momentum,
\be
\vk & = & 2 J \bp \sin(k_1) \\ \vdots \\ \sin(k_d)\ep.
\ee
This specifies the Boltzmann equation for the moment; we  study a specific collision functional in Sec.~\ref{sec:scattering} and thereafter.
The semiclassical momentum distribution
function $ \fk({\bf r},t)$  allows us to calculate several observables. In particular, the particle density $n$, kinetic energy density $e$,
current ${\bf j}$ and heat current ${\bf h}$ (or kinetic energy current) are obtained from the following formulas:
\be
n({\bf r},t) & = & \int {d\bf k}\, \fk({\bf r},t) /(2\pi)^d \n
e({\bf r},t) & = & \int {d\bf k}\, \ek \fk({\bf r},t) /(2\pi)^d \\
{\bf j}({\bf r},t) & = & \int {d\bf k}\, \vk \fk({\bf r},t) /(2\pi)^d \n
{\bf h}({\bf r},t) & = & \int {d\bf k}\, \ek \vk \fk({\bf r},t) /(2\pi)^d \nonumber
\ee
Here and in the following, the momentum integral is taken over the Brillouin zone $B^d \sim [-\pi,\pi]^d$ with periodic boundary conditions,
and the integration measure is $d\bk = \prod_{i=1}^d d\bk^i$. For notational convenience, we sometimes suppress the dependence
of the distribution function on ${\bf r}$ and $t$ and write $\fk$ for $\fk({\bf r},t)$.
Our goal is to find an approximate solution for $\fk$.

Let us furthermore specify the regime of interest. We assume that our distribution function is close to constant in the Brillouin zone, 
\be
&\fk({\bf r},t) =n({\bf r},t) + \delta \fk({\bf r},t),&\label{eq:ansatz}\\[.3cm]
&{\rm max }_{\;\bk \in B^d} \;|\delta \fk({\bf r},t)| \;\ll \; 1.&\nonumber
\ee
Formally, the expression in the second line above will be our perturbative expansion parameter. 
As a consequence, 
\be
e \ll 4dJ,
\ee
and hence the kinetic energy is close to the center of the band, which corresponds to maximal entropy. 
Therefore, our expansion can be thought of as a high-entropy expansion.
Typical distribution functions in this regime are sketched in the right plot  in Fig.~\ref{fig:linearize}, while typical distributions at low entropy in the linear response regime are shown at the left.
The plot expresses the fact that linearization at a high entropy
can be valid even when driving fields are large and the response is nonlinear.

The first step in our approach is the linearization of the Boltzmann equation in the deviations $\delta f$ from equilibrium. Using Eq.~(\ref{eq:ansatz}), we find
\be
I^{c}_\bk[n + \delta f] = I^{c}_\bk[n] +  \int_{\bf k'} \frac{d\bk}{(2\pi)^d} M(n)_{\bf kk'} \delta f_{\bf k'} \;+ \;{\cal O}(\delta f^2) \label{eq:lin_I}
\ee
where $M(n)$ is the matrix of the linearized collision functional, acting in momentum space. 
We stress that $M$ is a nontrivial operator, even in the limit of maximal entropy, which  becomes clear in 
Sec.~\ref{sec:scattering} where we specialize our approach to a specific collision integral. 
As the constant distribution function $n({\bf r})$ is an equilibrium distribution at maximal entropy for any collision functional, we have $I^c_\bk[n({\bf r})]=0$.
Furthermore, the matrix of the linearized collision functional has two important properties: it conserves the particle number and kinetic energy:
\be
\int_{\bf k'}d{\bf k}'\, M_{\bf kk'} n = 0,\quad \int_{\bf k'}d{\bf k}'\,M_{\bf kk'} \epsilon_{\bf k'} = 0.
\ee
Because of the first identity, the linearized Boltzmann equation
reads
\be
    \partial_t f + \mathbf{v}_\bk \cdot \nabla_\br f + \mathbf{F} \cdot \nabla_\bk f = -M(n({\bf r},t))\cdot f \label{eq:linBoltz},
\ee
where $"\cdot"$ denotes the matrix product.
Note that due to the dependence of $M$ on the local density $n$,
the above equation is still a non-linear equation in $f$. This equation  is the starting point of our further investigation. 

Since we have to deal with many momentum integrals and convolutions in this paper, it turns out to
be convenient to introduce some notation. We define a scalar product in the space of real, periodic functions of $\bf k$ in the Brillouin zone $B^d$ as

\be
\langle f_{\bf k}|g_{\bf k}\rangle := \frac{1}{(2\pi)^d}\int d{\bf k}\, f_{\bf k}\,g_{\bf k} \label{eq:BOdefScalar}
\ee
Our goal is to truncate the linearized Boltzmann equation to a minimal set of modes that captures
the essential physics. 
Focusing on the nonequilibrium transport of mass and energy, we write the following minimal ansatz:
\be
\fk({\bf r},t) = n({\bf r},t) + \frac{e({\bf r},t)}{2J^2d}\ek + \frac{{\bf j}({\bf r},t)}{2 J^2}\vk + \frac{{\bf h}({\bf r},t)}{6 J^4}\ek\vk
\label{eq:ansBOfk} \n
\ee 
It involves the variational parameters $n({\bf r},t)$, $e({\bf r},t)$, ${\bf j}({\bf r},t)$, and ${\bf h}({\bf r},t)$, which are in fact functions 
of the spatial coordinate ${\bf r}$ and time $t$. One should stress that the above ansatz for the nonequilibrium distribution function is very common in a linear response setup
where these parameters are assumed to be constant~\cite{ziman}, giving rise to nontrivial thermoelectric effects. Yet, our approach will allow them to vary in time, giving rise to
non-linear response effects.

Using the above scalar product and the above ansatz, it is straightforward to verify that 
\be
n  & =  \langle 1| \fk\rangle,   & e  =  \langle \ek |\fk\rangle,\n 
{\bf j} &  =  \langle \vk |\fk\rangle, & {\bf h}  =  \langle \ek\vk |\fk\rangle,
\ee
Hence, the coefficients $n$, $e$, ${\bf j}$, and ${\bf h}$ are exactly the particle density, kinetic energy, mass, and kinetic energy currents, respectively. 
Above, we have made use of the following integral identities:
\be
&\int \frac{d{\bf k}}{(2\pi)^d}\,\ek^2   =   2J^2d, \quad \int \frac{d{\bf k}}{(2\pi)^d}\, {\bf v}^i_{\bf k} {\bf v}^j_{\bf k}  =  2 J^2 \delta_{ij},&\label{eq:intIdentities} \\
&\int \frac{d{\bf k}}{(2\pi)^d}\, \ek^2{\bf v}^i_{\bf k}{\bf v}^j_{\bf k}   =   6 J^4 \delta_{ij}&\nonumber
\ee
which explains the choice of numerical prefactors in Eq.~(\ref{eq:ansBOfk}). 

Is is worth mentioning that the conventional variational approach would follow a different strategy from this point on, which we 
sketch in the next section. Here, 
we project the linearized Boltzmann equation,~(\ref{eq:linBoltz}), onto the mode functions of our variational ansatz:
\be
\langle 1|\left(\partial_t + \vk \nabla_{\bf r} + {\bf F}\nabla_{\bf k}\right) \fk \rangle & = & 0 \label{eq:bracketCont}\\
 \langle \ek|\left(\partial_t + \vk \nabla_{\bf r} + {\bf F}\nabla_{\bf k}\right) \fk \rangle & = & 0\n
 \langle \vk|\left(\partial_t + \vk \nabla_{\bf r} + {\bf F}\nabla_{\bf k}\right) \fk \rangle & = & -\langle \vk|M|\fk\rangle\n
 \langle \ek\vk|\left(\partial_t + \vk \nabla_{\bf r} + {\bf F}\nabla_{\bf k}\right) \fk \rangle & = & -\langle \ek\vk|M|\fk\rangle\nonumber
\ee
Above, $|1\rangle$ denotes the constant function in momentum space. We have
used that the scattering terms vanish for the particle number and energy modes, i.e.,
$\langle 1 |M|\fk \rangle = 0$ and $\langle \ek|M|\fk\rangle = 0$, as discussed above. 
Using the ansatz Eq.~(\ref{eq:ansBOfk}), Eq.~(\ref{eq:intIdentities}) and the orthogonality of the different momentum modes allows us
to reformulate these equations. In fact, we arrive at a set of coupled differential equations for the coefficients 
$n$, $e$, ${\bf j}$, and ${\bf h}$ :
\be
\dot{n} + \nabla_{\bf r}{\bf j} & = & 0 \label{eq:BOseteqns}\\
\dot{e} + \nabla_{\bf r}{\bf h} - {\bf F}{\bf j} & = & 0 \n
\dot{{\bf j}} + 2J^2\, \nabla_{\bf r}n + {\bf F}\,e & = & - M_{11} {\bf j} -M_{12} {\bf h}\n
\dot{{\bf h}} + \frac{3}{d} J^2\,\nabla_{\bf r}e  & = &- M_{21} {\bf j} - M_{22}  {\bf h}\nonumber
\ee
This calculation is presented in Appendix~\ref{sec:app1}. Equation~(\ref{eq:BOseteqns}) is one of the central results of this paper.
Above, we used the following $d\times d$ matrices: 
\be
 M_{11} =  \frac{1}{2J^2}\langle \vk|M|\vkt\rangle, \quad M_{12}  =  \frac{1}{6J^4}\langle \vk|M|\ek\vkt\rangle, \n
 M_{21}  =  \frac{1}{2J^2}\langle \ek\vk|M|\vkt\rangle, \quad M_{22}  =  \frac{1}{6J^4}\langle \vk|M|\ek\vkt\rangle  \nonumber
\ee
The first two Eqs.~(\ref{eq:BOseteqns}) are nothing  
but the continuity equations for the particle and kinetic energy density, respectively.
The source term ${\bf Fj}$ in the kinetic energy continuity equation corresponds to Joule heating.
The third and fourth equations, in contrast, describe the dynamics of the particle and heat current, respectively. In contrast to the
previous two modes, these modes are damped by the scattering matrix elements.
In the remainder of this article, we study different limits of these equations. To begin with, we  show
that they describe hydrodynamics as a limiting case. 

\section{The hydrodynamic limit}
\label{sec:hydrodynamics}

First, we demonstrate that our ansatz captures hydrodynamics at high entropies as a limiting case. 
In the hydrodynamic limit, the Boltzmann equation is reduced to coupled equations for the particle density $n({\bf r},t)$ and kinetic energy density $e({\bf r},t)$.
The number of coupled hydrodynamic equations is determined by the number of conservation laws (here particle number and energy).
We consider the case where momentum is not conserved, e.g., due to Umklapp scattering processes.
The hydrodynamic limit amounts to expressing ${\bf j}$ and ${\bf h}$ in terms of gradients of $n$ and $e$.
As the currents passively follow these gradients, their own  dynamics can be thought of as overdamped (no retardation). We 
show now that this picture is indeed correct: in setting the time derivatives of ${\bf j}$ and ${\bf h}$ to $0$,
we derive the hydrodynamic equations in the high-temperature (high-entropy) limit.

We begin by setting the time derivatives of ${\bf j}$ and ${\bf h}$ in Eq.~(\ref{eq:BOseteqns}) to $0$.
The particle and heat currents in the diffusive limit are hence a solution to the inverse problem,
\be
\bp 2J^2\, \nabla_{\bf r}n + {\bf F}\,e \\ \frac{3}{d} J^2\,\nabla_{\bf r}e \ep =-\begin{pmatrix} M_{11} && M_{12} \\ M_{21} && M_{22} \end{pmatrix} \cdot \bp {\bf j} \\ {\bf h} \ep \label{eq:inv1}
\ee
these equations can be solved for ${\bf j}$ and ${\bf h}$ and , in combination with the continuity equations, form a closed set of equations
for $n$ and $e$. 

We  now sketch how to produce the same result using the conventional variational approach. 
There, the diffusive currents are calculated from the linearized Boltzmann equation,~(\ref{eq:linBoltz}), in 
decomposing $f = f^0(n,e) + \delta f$, where $\delta f$ is assumed to be a small deviation from equilibrium. Neglecting the time derivative 
and $\delta f$ on the left-hand side yields relation structurally similar to Eq.~(\ref{eq:inv1}):
\be
(\vk \nabla_{\bf r} + {\bf F} \nabla_{\bf k}) f^0 = - M \cdot \delta f  \label{eq:inv2}
\ee
We use our variational ansatz, (\ref{eq:ansBOfk}), for $f$ and decompose it into $f^0(n,e) = n + e \, \ek /(2J^2d)$ 
and $\delta f = f - f^0 = {\bf j} \, \vk / (2 J^2) + {\bf h}\, \ek \vk / (6 J^4)$. One can easily check that $f^0$ approximates
the Fermi function to leading order in $e$, see also Ref.~\cite{gravity}.
Using this decomposition, Eq.~(\ref{eq:inv2}) becomes
\be
& & (\nabla_{\bf r}n + e {\bf F})\vk + \nabla_{\bf r}e \,\ek \vk \\[.2cm] &=& - M \cdot \left(\frac{{\bf j}({\bf r},t)}{2 J^2}\vk + \frac{{\bf h}({\bf r},t)}{6 J^4}\ek\vk\right). \nonumber
\ee
This equation gives exactly Eq.~(\ref{eq:inv1}) if we project it onto the modes  $\vk$ and $\ek \vk$ and use Eq.~(\ref{eq:intIdentities}). 
We have just demonstrated that using the conventional variational approach at high entropies yields the same result as obtained from our new method, 
where we manually set the time derivatives of ${\bf j}$ and ${\bf h}$ to $0$. For more details on hydrodynamics at high energies in this setup, see also~\cite{gravity}.
We  show below that our method is not limited to the diffusive regime but models  nonlinear transport as well.

\section{Weakly interacting Fermions and incoherent bosons}
\label{sec:scattering}

For the remainder of the paper, we  concentrate on interparticle scattering. We consider fermions and incoherent bosons. 
We  study the semiclassical Boltzmann equation for two-particle scattering processes due to its relevance for ultracold atoms in optical lattices.
This allows us to calculate the scattering elements $M_{ij}$ that fully specify Eq.~(\ref{eq:BOseteqns}).

The fermionic Boltzmann equation can be derived from the Hubbard model in the presence of a linear potential, 
\be
H_F=-J \sum_{\langle ij \rangle, \sigma} c^\dagger_{i\sigma} c_{j\sigma} + U \sum_i n_{i\uparrow} n_{i \downarrow} + {\bf g} \sum_i {\bf r}_i n_i
\ee
Similarly, a bosonic Boltzmann equation can be derived from the Bose-Hubbard model,
\be
H_B=-J \sum_{\langle ij \rangle} b^\dagger_{i} b_{j} + U \sum_i n_{i} (n_{i} - 1) + {\bf g} \sum_i {\bf r}_i n_i
\ee
In second order perturbation theory in $U/J$, the resulting Boltzmann equations have the following collision integral,
\be
\label{eq:fullCollision}
&&I^c_\bk[f]  =  2\pi U^2 \int \frac{d{\bk_1}}{(2\pi)^d}\frac{d{\bk_2}}{(2\pi)^d}\frac{d{\bk_3}}{(2\pi)^d} \\[.1cm]
&&	     \times \left[\fk f_{\bk_1}\bar{f}_{\bk 2}\bar{f}_{\bk 3} -  \bar{f}_{\bk}\bar{f}_{\bk_1} f_{\bk 2} f_{\bk 3}\right] \nonumber \\[.1cm]
&&	     \times  \delta({\bk+\bk_1-\bk_2-\bk_3} {\rm \,mod\, } {\bf G}) \, \delta(\epsilon_{\bf k} + \epsilon_{\bk_1} - \epsilon_{\bk_2} - \epsilon_{\bk_3}) \nonumber.
\ee
Fermionic and bosonic statistics enter via the following factors, describing Pauli blocking and Bose enhancement:
\be
\bar{f}_{\bk_i} &=& \begin{cases} (1+  f_{\bk_i}) & \textrm {for bosons}     \\
(1-  f_{\bk_i}) & \textrm {for fermions}
 \end{cases}
\ee
In the case of fermions, we assumed a homogeneous mixture of spins for simplicity, and hence the distribution function $f_{\bf k} = f_{\bf k \uparrow} =f_{\bf k\downarrow} $
describes one spin component.

A systematic derivation of the Boltzmann equation from 
an underlying quantum Hamiltonian can be found, e.g., in Refs.~\cite{rammer,kamenev:book,mahan,zaremba}. 
This derivation is based on several approximations including weak interactions and slowly varying 
gradients and external potentials. In order to justify the real-space basis rather than a description 
in terms of Wannier-Stark states, we have to assume that
the external potential $V({\bf r})={\bf g \cdot r}$ is weak. The semiclassical model still captures the physics 
of Bloch oscillations due to the periodicity of the kinetic energy $\ek$ and group velocity $\vk$ in $\bf k$.

Our theory is based on the Boltzmann equation,~(\ref{eq:fullCollision}). We 
do not aim to describe anything beyond its range of validity. 
The Boltzmann equation can, of course, only capture parts of  the rich physics described by the Hubbard model.
Many physical effects that involve strong interactions and quantum coherence are lost.
In particular, we treat bosons essentially as classical particles with Bose statistics, and hence in this case,
our approach can only describe an incoherent, highly entropic Bose gases. For fermions, we cannot describe
any ordered state.  The fermionic Boltzmann equation was quantitatively tested as a means to model 
ultracold atoms in optical lattices in Ref.~\cite{expansion}, where good agreement was found. 

In the collision integral,~(\ref{eq:fullCollision}), we took 
Umklapp scattering into account, which is crucial.
Umklapp scattering processes are scattering events that violate momentum conservation - 
but satisfy momentum conservation modulo reciprocal lattice vectors ${\bf G}$. 
This can be thought of as a momentum transfer to the lattice. Umklapp scattering breaks translational 
invariance and favors equilibration into the frame of reference set by the lattice.
This is necessary to have finite conductivities and scattering matrix elements in a clean and defect-free lattice. 

We now use the standard methodology 
of computing matrix elements of the linearized collision integral~\cite{ziman}. To simplify the notation, we define 
\be
\bphi^i_\bk = \begin{cases} {\bf v}^i_{\bf k} & 1 \leq i \leq d \\ \ek  {\bf v}^{i-d}_{\bf k}  & d < i \leq 2d. \end{cases}
\ee
We expand the collision integral around its maximum entropy equilibrium solution $n({\bf r},t)$. From Eq.~(\ref{eq:lin_I}) we derive the following formula for the matrix elements:
\be
\langle \bphi^i_\bk  |M| \bphi^j_\bk \rangle & = & \frac{\partial}{\partial \varepsilon} \biggr\rvert_0 \langle \bphi^i_\bk  \; | \;  I^c_{\bk}[n +\varepsilon \, \bphi^j_\bk ] \, \rangle
\ee
The resulting matrix elements
are
\be
&&\langle \bphi^i_\bk |M|\bphi^j_\bk\rangle  =  \frac{2\pi U^2}{4} n(1\pm n)\int \frac{d{\bk_0}}{(2\pi)^d}\frac{d\bk_1}{(2\pi)^d}\frac{d{\bk_2}}{(2\pi)^d}\frac{d{\bk_3}}{(2\pi)^d} \label{eq:matrixElements}\nonumber \\[.1cm]
&&	     \times  \left( \bphi^i_{\bk_0} + \bphi^i_{\bk_1} - \bphi^i_{\bk_2} - \bphi^i_{\bk_3}\right) \left( \bphi^j_{\bk_0} + \bphi^j_{\bk_1} - \bphi^j_{\bk_2} - \bphi^j_{\bk_3}\right) \nonumber \\[.1cm]
&&	     \times  \delta({\bk_0+\bk_1-\bk_2-\bk_3} {\rm \,mod\, } {\bf G}) \, \delta(\epsilon_{\bk_0} + \epsilon_{\bk_1} - \epsilon_{\bk_2} - \epsilon_{\bk_3}). \n
\ee
The fermionic scattering rate is proportional to $n(1-n)$, which reflects particle-hole symmetry. 
The bosonic one is proportional to $n(1+n)$, which captures Bose enhancement. 

The resulting integrals can be computed numerically. Below we calculate them analytically for a one-dimensional discrete Boltzmann equation.
Our analysis implies that certain off-diagonal scattering matrix elements are strictly $0$:
\be
\langle \ek{\bf v}^{i}_{\bf k}|M|{\bf v}^{j}_{\bf k}\rangle = \langle {\bf v}^{i}_{\bf k} |M|\ek{\bf v}^{j}_{\bf k}\rangle = 0
\ee
for all $i,j\in\{1,\cdots,d\}$.
This can be seen by translating all momenta in the integral in Eq.~(\ref{eq:matrixElements}) by $\pi$. This transformation
leaves the delta-constraints and the heat current modes $\ek\vk$ invariant. In contrast, the particle current mode changes sign, $\vk \rightarrow -\vk$,
which is why the integral is $0$. This implies that thermoelectric effects vanish at maximal entropy, which is usually not the case at low temperatures.

The second property is that, by symmetry, the current-current scattering matrix is of the  shape (shown here for $d=3$)
\be
\frac{1}{2J^2}\langle \vk|M|\vk\rangle = 2\bp \tau^{-1}_d && \tau^{-1}_o && \tau^{-1}_o \\ \tau^{-1}_o && \tau^{-1}_d && \tau^{-1}_o \\ \tau^{-1}_o && \tau^{-1}_o && \tau^{-1}_d\ep \label{eq:structScatt},
\ee
i.e., the matrix decomposes into diagonal scattering rates $\tau_d^{-1}$ and off-diagonal scattering rates $\tau_o^{-1}$ which are all identical.
These two properties are important in the following analysis.

In the following, we  apply our approach to two nontrivial examples and derive novel results: the damping of Bloch oscillations
due to interactions in a spatially homogeneous system and the expansion of a cloud of fermions or incoherent bosons in a tilted lattice subject to Bloch oscillations,
which we term stroboscopic diffusion.

\section{Damping of Bloch oscillations}
\label{sec:blochoscillations}

We  now use our approach to model the damping of Bloch oscillations in the Hubbard model. We assume that 
the conditions for its semiclassical treatment in terms of the Boltzmann equation apply.

Bloch oscillations emerge in lattice systems that are exposed to an additional linear potential. 
Without interactions, they are captured already by the semiclassical equations of motion
$\dot{\bf r} = \vk$ and $\dot{\bk} = {\bf F}$. These equations can easily be integrated, yielding a linearly growing momentum $\bk = \bk_0 + {\bf F}t$.
The periodic group velocity ${\bf v}_{\bk(t)}$ imposes an oscillatory motion of the position space variable ${\bf r}(t)$.
It is an interesting question to ask how this behavior changes in the presence of inter-particle interactions.

Bloch oscillations belonged to the pioneering observations with ultracold atoms in optical lattices~\cite{kasevich,salomon}. 
They were first predicted for electrons in periodic potentials~\cite{bloch0}.
The difficulty in observing Bloch oscillations in regular solids lies in the fast time scale of scattering on impurities, phonons, lattice defects or other electrons,
relative to the driving strength.
Besides ultracold atoms, there
are more quantum systems that show these oscillations, among which are semiconductor-superlattices~\cite{waschke}, 
and optical waveguide arrays~\cite{BOwaveguides}.

The damping of Bloch oscillations has been observed in experiment with ultracold atoms~\cite{gustavsson:BlochOsc},
and it has been studied theoretically in many physical realizations~\cite{schecter,freericks:DMFT,freericks:DMFT2,prelovsek,ponomarev:dephasing,kolovsky2010,
werner:BO, capone:BO, freericksNew,buchleitner}. For various fermionic lattice models, the damping of Bloch oscillations
has been studied numerically using dynamical mean field theory~\cite{freericks:DMFT,freericks:DMFT2,werner:BO, capone:BO}. 

At this point, let us address an important aspect of the driven Hubbard model
and the corresponding Boltzmann equation. Both models describe thermally isolated systems: all Joule heating that is generated by the current remains in the system. 
Also, ultracold atoms in optical lattices are thermally isolated. Therefore, they realize the Hubbard model better than electrons in solids,
which are coupled to a bath. 
When being driven by a constant force, the fermions in a lattice monotonously heat up.
The system in the long-time limit 
is therefore in a maximum entropy state, characterized by a flat momentum distribution and a vanishing current. 

Eckstein and Werner simulated  damping of Bloch oscillations and  heating in the fermionic Hubbard model
by an electric field numerically~\cite{werner:BO}. 
For weak interactions, they found the current to oscillate and with an exponentially decaying amplitude.
For stronger interactions, these oscillations were found to become overdamped, and the current was found to decay exponentially. 
Below, we  present an analytic theory of the damping of Bloch oscillations, showing the same phenomena.

Buchleitner and Kolovsky~\cite{buchleitner} studied bosonic atoms in a homogeneous lattice that were initially
in the superfluid phase, using a lattice Gross-Pitaevskii equation. They showed the irreversible decay of Bloch oscillations by interactions. 
This approach is different from ours: for bosons, we consider a fully incoherent system at high entropy.  This allows us to use the Boltzmann equation.

In this section we consider a homogeneous system. Homogeneity implies the absence of all
spatial gradients in Eqs.~(\ref{eq:BOseteqns}). Due to the anisotropy of the lattice, the direction of the constant force also matters. 
As in other studies on this topic~\cite{werner:BO,freericks:DMFT}, 
we  consider a setup in which the force points into the lattice diagonal,
\be
{\bf F} = F\, {\bf 1} = \underbrace{(F,F,...,F)^T}_{d}.
\ee
As discussed above, the absence of thermoelectric effects close to maximal entropy leads to a decoupling 
of the velocity and heat current modes via scattering. Due to the structure~(\ref{eq:structScatt}) of the current-current scattering matrix, 
it is enough to parametrize ${\bf j} = j\,{\bf 1}$ and study the two coupled equations,
\be
\dot{e} - F\,j & = & 0 \label{eq:BOCoupledHomo} \\
\dot{j} +F \,e & = & -2 \tau^{-1} \,j \nonumber
\ee
where
\be
\tau^{-1} = \tau_d^{-1} + (d-1)\tau_o^{-1}
\ee
Equations~(\ref{eq:BOCoupledHomo}) can be combined into the following second-order differential equation,
\be
\ddot{j} = F^2\, j - 2\tau^{-1}\,\dot{j} \label{eq:BOHO}
\ee
This is nothing but the equation of the classical damped harmonic oscillator, which is solved by
\be
j(t)& =& \,e^{-t/\tau}\,\left[A\,\exp\left(t\,{\sqrt{\tau^{-2}-F^2}}\right)\right.\label{eq:vtAB} \\
& &\left. + B\,\exp\left(-t\,{\sqrt{\tau^{-2}-F^2}}\right)\right] \nonumber
\ee
for arbitrary constants $A$ and $B$.
We have thus mapped the damping of the kinetic energy in the Hubbard model in the presence of a constant
force to the classical harmonic oscillator. Note that the oscillator frequency (which is the Bloch frequency $\omega_B$ for $F > \tau^{-1}$) depends on the scattering rate,
\be
w_B = \sqrt{F^2 - \tau^{-2}}
\ee
and that it goes to $0$ at $F = \tau^{-1}$. This critical value of $F$ separates overdamped from underdamped Bloch oscillations. 

In the \textit{overdamped} limit, $F\tau\ll1$, we can approximate 
the square root in the full solution Eq. (\ref{eq:vtAB}) as
\be
\sqrt{1/\tau^{2}-F^2}\; \approx \;\tau^{-1} \left( 1 - \frac{1}{2}(\tau F)^2 \right)
\ee
Therefore, the velocity and hence also the kinetic energy mode decay according to
\be
e(t) \approx e_0\, \exp \left(-\frac{1}{2}\,t\,\tau\,F^2\right), \quad F\tau\ll1\label{eq:BO-over}
\ee
in this limit.
The damping rate is hence given by $\tau F^2/2$, which can also be derived from the conventional transport theory for an isolated system. 
In the opposite case of strong Bloch oscillations, 
$F\,\tau\gg 1$, the square roots in Eq. (\ref{eq:vtAB}) become negative. 
Equation~(\ref{eq:vtAB}) can be approximated as
\be
e(t)\approx e_0\, e^{-t/\tau}\,\cos(\omega_B t), \quad F\,\tau\gg 1 \label{eq:BO-under}
\ee
Most prominently, the oscillations decay exponentially at the rate $\tau^{-1}$.
Note that this result can not be obtained from conventional transport theory, as the 
response of the induced currents to the driving field is highly nonlinear (namely oscillatory) in this limit. 
Finally, in the marginal case of $F\tau=1$, Bloch oscillations get critically damped. For the initial condition  $e(0)=e_0$ and
$\dot{e}(0)=0$, the dynamics of the kinetic energy mode is given by
\be
e(t) = \left( e_0 + e_0 t/\tau\right)e^{-t/\tau}, \quad F \tau=1 \label{eq:BO-marginal}
\ee 
which contains corrections to a purely exponential decay of the energy mode. Note that this limit is highly non-perturbative
in the ratio $\tau F$.

\subsection*{Comparison with Boltzmann simulations}

We compare our analytic results for the damping of Bloch oscillations to a simulation of the Boltzmann equation. 
We simulate a discrete version of the one-dimensional Boltzmann
equation with a collision integral,~(\ref{eq:fullCollision}), as a toy model. This is a computationally tractable model that also allows us to calculate the scattering matrix elements analytically. 
Note that it does not capture the physics of the one-dimensional Hubbard model due to the integrability of the latter. 

Abbreviating $f_i = f_{k_i}$, the equation reads
\be
\dot{f}_{i} + F\frac{f_{i+1}-f_{i-1}}{2\Delta k} = I[f]_{k_i} \label{eq:1DBoltz_1}
\ee
This involves the discrete collision integral
\be
I[f]_{k} & = & -\frac{U^2}{J}\frac{1}{N}\sum_{k_1,k_2,k_3}\; \delta \left(\epsilon_k + \epsilon_{k_1}-\epsilon_{k_2}-\epsilon_{k_3}\right)\n
          & \times&   \left(f_k f_{k_1} \bar{f}_{k_2} \bar{f}_{k_3} - \bar{f}_{k} \bar{f}_{k_1}f_{k_2}f_{k_3}\right) \label{eq:1DBoltz_2} \\
	  & \times & \delta(k+k_2-k_2-k_3\,{\rm mod}\,\pi)\nonumber.
\ee
We simulated a fermionic system:
\be
\bar{f}_{k_i}=(1-f_{k_i})
\ee
The momenta $k_1, k_2$ and $k_3$ in the sum run over the $N$ discrete values $\{ -1 + \frac{1}{N},-1 + \frac{3}{N}, ... , -1 + \frac{2N-1}{N} \}\times \pi$.
The delta symbols denote discrete Kronecker delta constraints. 

As in our previous studies~\cite{expansion,gravity}, the presence of Umklapp processes is crucial to ensure equilibration to the fixed frame of reference of the lattice. 
The distribution function $\fk$ was initially prepared as a Fermi function at $T=J$, a typical temperature for current experiments with ultracold fermionic atoms. 

For our discrete one-dimensional Boltzmann equation, the matrix elements that lead to a relaxation of the current-current mode~(\ref{eq:matrixElements}) can be calculated analytically. 
The calculation is shown in Appendix \ref{sec:AppCurrCurr}. 
For a large number of discrete momenta $N$, the matrix element is 
\be
\tau^{-1} = \langle v_k|M|v_k\rangle\;\approx\; 4\,n(1-n)\,U^2\,J
\ee
We considered the fermionic case; the bosonic case yields $4\,n(1+n)\,U^2\,J$ (see Appendix~\ref{sec:AppCurrCurr}).
This formula for the scattering rate concludes our analytic result.

\begin{figure}
 \begin{center}
  \includegraphics[width=\linewidth]{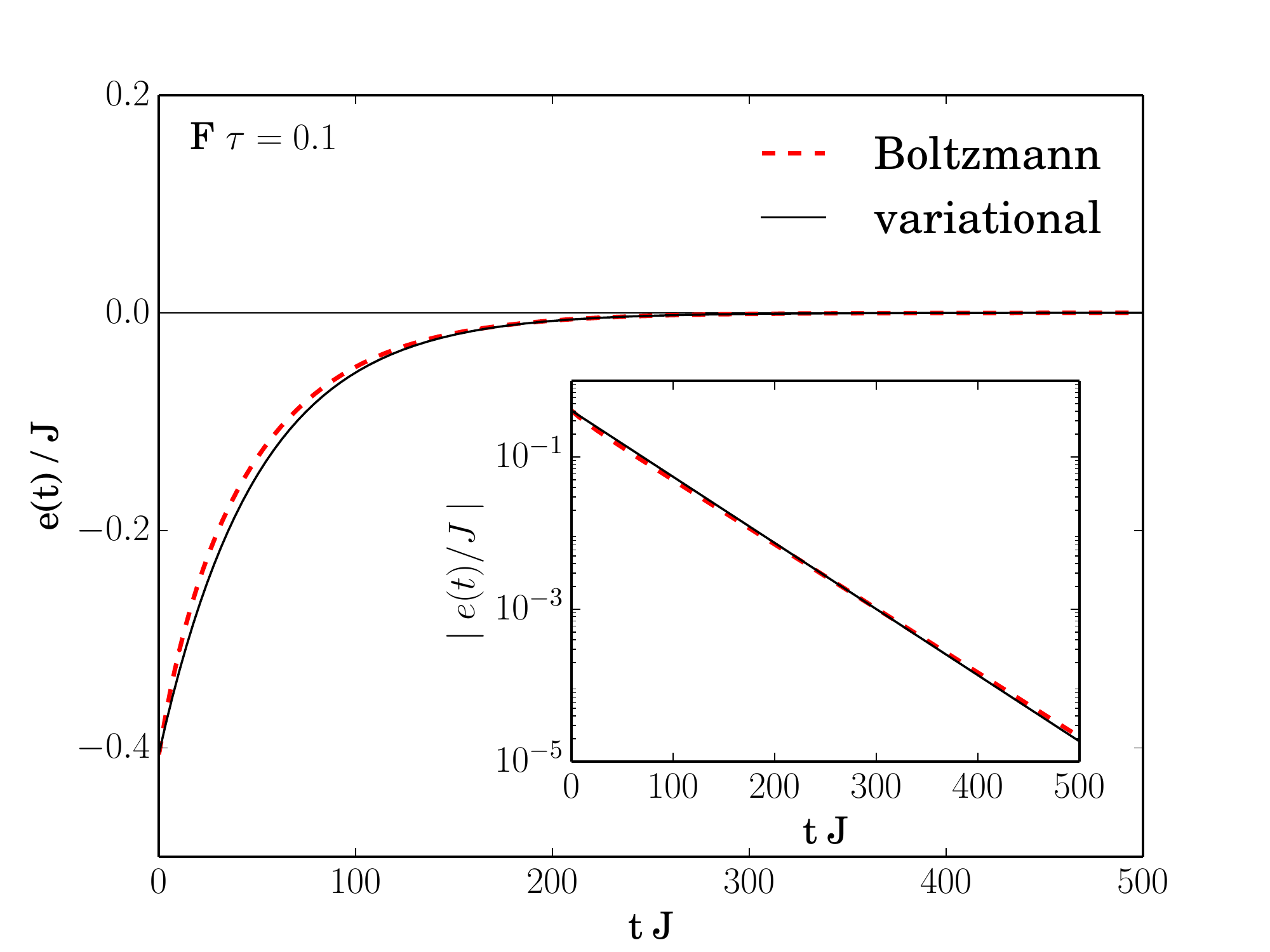}
  \caption{\CO \textit{Overdamped} Bloch oscillations of the kinetic energy at filling $1/2$. We compare a full numerical simulation [dashed, red curve]
of the Boltzmann equation (\ref{eq:1DBoltz_1}) with our analytic result from Eq. (\ref{eq:BO-over}) [solid, black curve].
  The parameters  are $U/J=4$ (yielding $\tau=0.25/J$) and $F=0.4 J$ such that $F\tau=0.1$.}  \label{fig:BOoverdamped}
 \end{center}
\end{figure}
  
\begin{figure}
 \begin{center}
  \includegraphics[width=\linewidth]{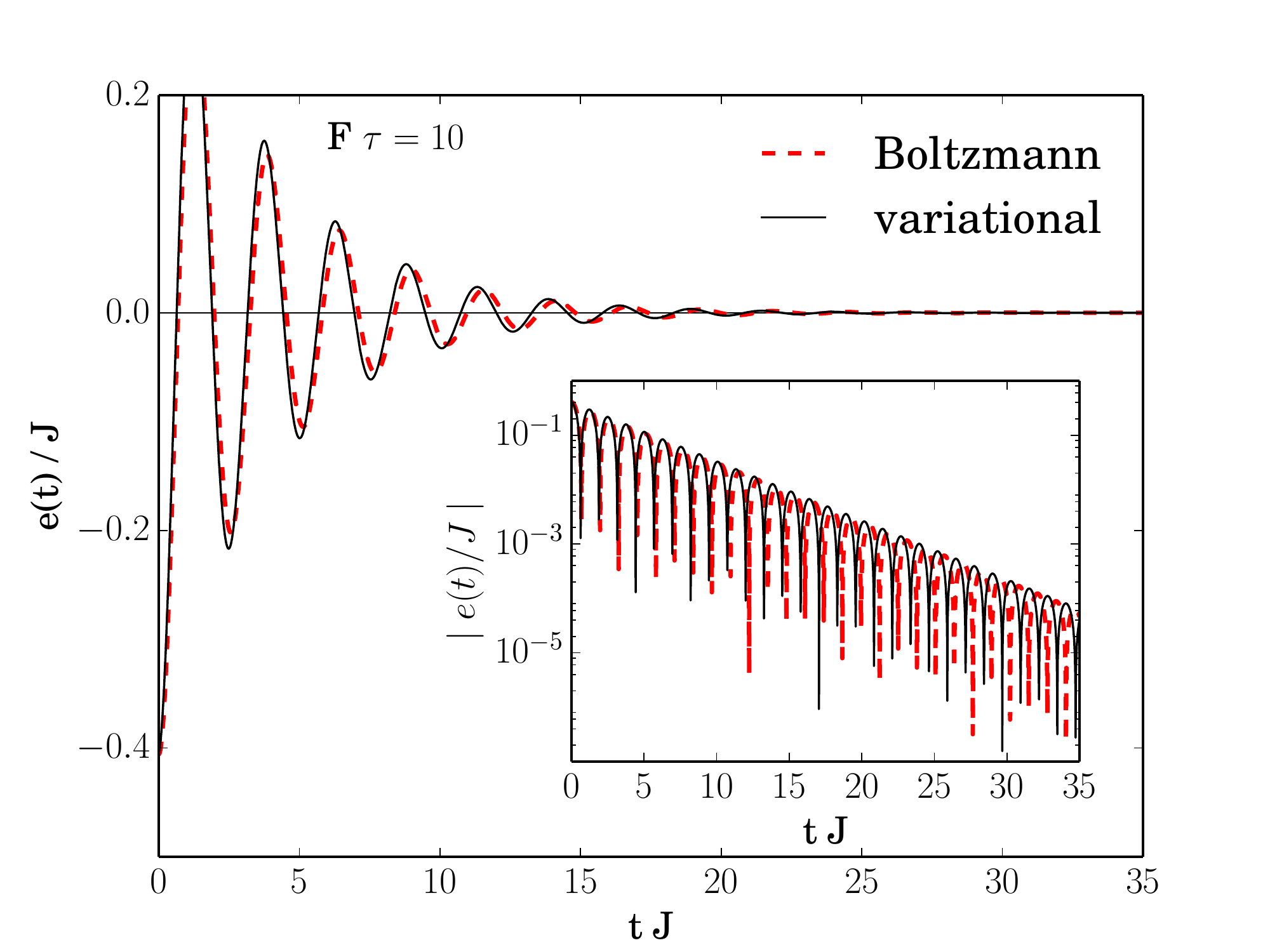}
  \caption{\CO \textit{Underdamped} Bloch oscillations of the kinetic energy at filling $1/2$. Dashed (red) curve: numerical simulation of the Boltzmann equation (\ref{eq:1DBoltz_1}) for 
  the parameters $U/J=1$ (yielding $\tau=4/J$) and $F=2.5 J$ such that $F\tau=10$ gives rise to the regime of weak damping. 
  Black solid curve: analytic result from Eq. (\ref{eq:BO-under}). }
  \label{fig:BOunderdamped}
 \end{center}
\end{figure}

\begin{figure}
 \begin{center}
  \includegraphics[width=\linewidth]{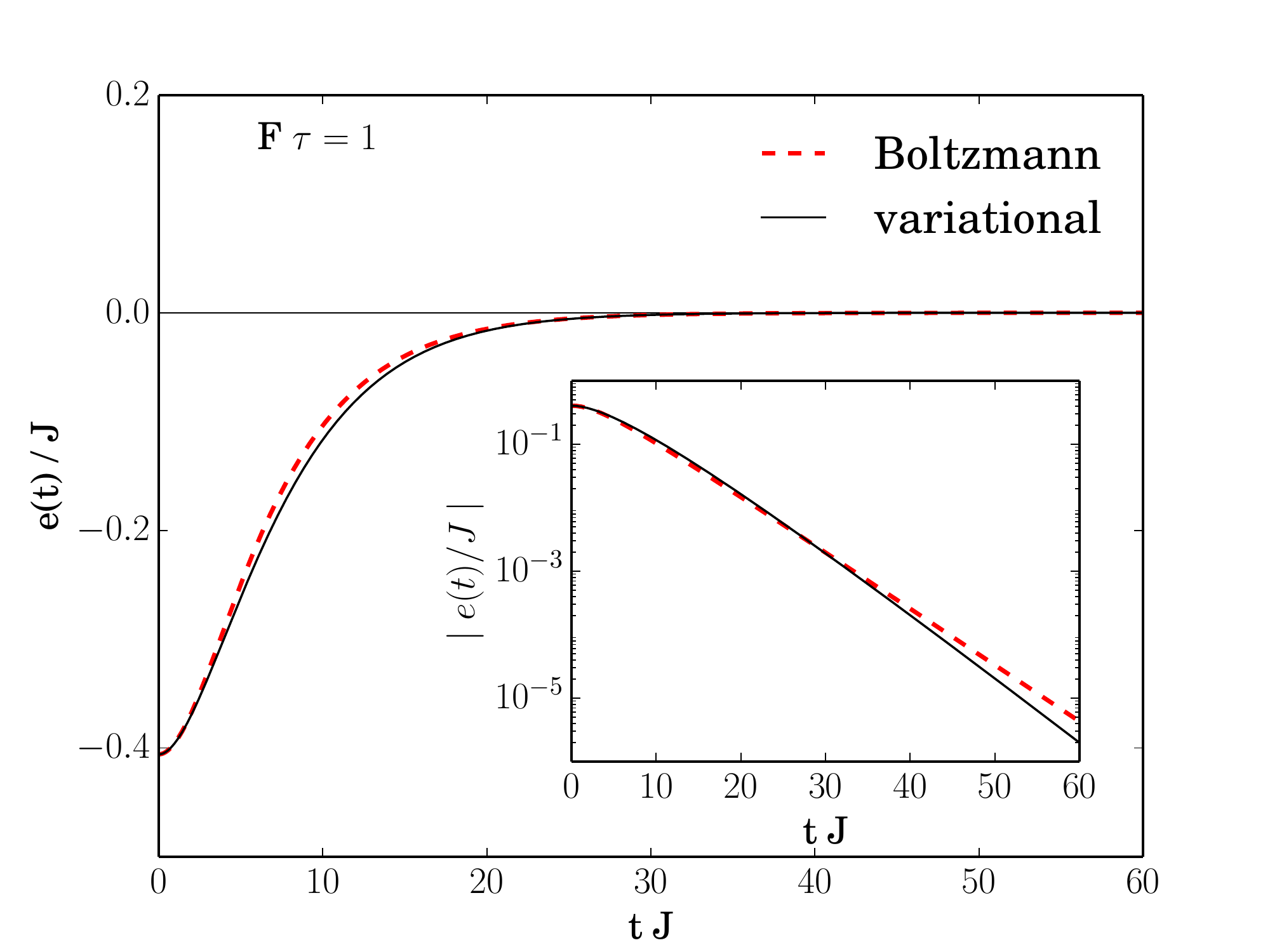}
  \caption{\CO \textit{Marginally} damped Bloch oscillations of the kinetic energy at filling $1/2$. Dashed (red) curve: numerical simulation of the Boltzmann equation (\ref{eq:1DBoltz_1}) for 
  the parameters $U/J=1$ (yielding $\tau=4/J$) and $F=0.25 J$ such that $F\tau=1$ gives rise to the marginal case. 
  Solid (black) curve: analytic result from Eq. (\ref{eq:BO-marginal}). }
  \label{fig:BOmarginallydamped}
 \end{center}
\end{figure}

We now compare our simple analytic theory with the numerical simulation of the discrete one-dimensional Boltzmann equations, ~(\ref{eq:1DBoltz_1}) and (\ref{eq:1DBoltz_2}).
We  consider the case $n=1/2$ for which we have $\tau \;=\; 4\, J/U^2$. 

Figures \ref{fig:BOoverdamped}, \ref{fig:BOunderdamped}, and \ref{fig:BOmarginallydamped} show the cases of overdamped, underdamped
and marginally damped Bloch oscillations, respectively. While the dashed (red) lines show the numerical simulations, Eqs.~(\ref{eq:1DBoltz_1}) and (\ref{eq:1DBoltz_2}), 
the  solid (black) lines show the analytic predictions given by Eqs. (\ref{eq:BO-over}), (\ref{eq:BO-under}), and (\ref{eq:BO-marginal}), respectively.
As for the initial conditions, we adjusted the initial kinetic energy $e_0$ to be the same for both methods, and chose $\dot{e}(0)=0$.  
As the analytic formulas depend only on the force $F$ and on the calculated damping rate $\tau^{-1}$, no fitting parameters were involved. 
Surprisingly, the analytic formulas describe the complex dynamics of the Boltzmann equation extremely well, despite  the
fact that they are based on a high-entropy expansion.

The weakly damped and overdamped regimes of Bloch oscillations have also been observed numerically for the Hubbard model by Eckstein and Werner~\cite{werner:BO}, using dynamical mean-field theory. 
Just as in our case, they found a transition from overdamped to underdamped Bloch oscillations. 
Our mapping of the problem to the harmonic oscillator equation might give an analytic explanation for the numerically observed transition.
While we are not able to capture the regime of strong interactions and strong drive, we are able to treat the case of weak interaction-induced damping. 
For a quantitative comparison, one would have to compute the corresponding scattering matrix elements in the limit of infinite spatial dimensions. 

We now proceed to a second application of our variational approach.

\section{An interacting, Bloch oscillating cloud in a tilted lattice}
\label{sec:stroboscopic}
\label{sec:BOinhom}

\begin{figure}[h]
 \begin{center}
  \includegraphics[width=.8\linewidth]{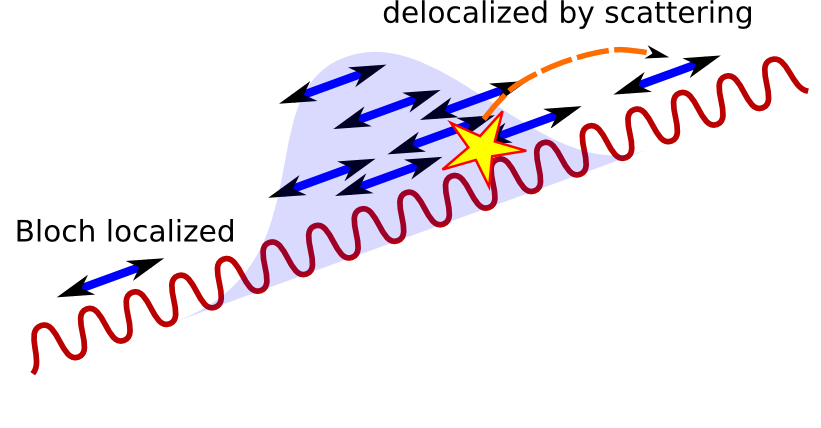}
  \caption{\CO Sketch of an interacting, Bloch oscillating cloud of atoms in a tilted lattice. Without interactions, individual particles are Bloch localized
   due to energy conservation and the bounded kinetic energy. This localization can, however, be lifted by inter-particle scattering. 
As the scattering rate increases with the particle density, we can expect fast diffusion in the bulk and slow diffusion in the tails of the cloud.
}
  \label{fig:BOsketch}
 \end{center}
\end{figure}

We now consider a generalization of the problem of damped Bloch oscillations in a spatially inhomogeneous situation.
The dynamics of a \textit{finite} cloud of bosons or fermions in a tilted lattice is an interesting problem with relevance for ultracold atoms in optical lattices.

In the regime where the potential energy difference between neighboring lattice sites is weaker than the scattering rate, $F \tau  \ll 1$ (note that we set the lattice constant and $\hbar$ to $1$), 
this problem can be studied in terms of coupled hydrodynamic
equations for particle and energy densities, which was done in Ref.~\cite{gravity}. Here, we are interested in the regime of strong Bloch oscillations,
$F \tau  \gg 1$, where the conventional hydrodynamic ansatz breaks down. For bosons, this problem was studied by Kolovsky\etal~\cite{kolovsky2010}, using the Gross-Pitaevskii framework,
and we  comment on the connection to our work below. Strong driving is a challenging problem which usually has to be studied using
the full Boltzmann equation, but it turns out that our variational method can be used to approximately solve this problem at high entropies. 

In this section, we consider a quantum gas at low densities $n$,  such that we approximate
\be
n(1-n) \; \approx \;  n(1+n) \; \approx \;  n.
\ee
Hence, both fermionic and bosonic particles essentially assume classical statistics and can be treated simultaneously (again we assume incoherent,
particle-like bosons). 
Note that even in this limit, quantum effects are still present and manifest themselves in the lattice dispersion relation,
allowing for Bloch oscillations. 

\subsection{The physical heuristic} Let us start by developing some physical intuition regarding this problem. First, note that
noninteracting particles in a tilted lattice are confined in position space. This is due to energy conservation and the fact that kinetic energies are bounded in a lattice.
As all individual particles are Bloch oscillating, the cloud's collective motion is periodic in units of  $\tau_B = 2\pi/F$.
In contrast, interacting particles can exchange energies by collisions, and therefore can explore a much wider range in position space.
Also, collisions between the particles break the periodicity of the cloud's collective motion. At weak interactions, we can 
expect the cloud to dominantly Bloch-oscillate, and slowly diffuse due to scattering events. 
We are interested
in the \textit{stroboscopic} motion of the cloud in units of $\tau_B$, separating the slow diffusive dynamics of the cloud  from it's fast Bloch oscillations.
Thus, we may ask about the corresponding diffusion constant $D^{strob}$ and its dependence on the local densities 
and system parameters. For the semiclassical limit of the Hubbard model at high energies and low densities, we have shown before that the scattering rate satisfies 
$\tau^{-1}(n) = \tau^{-1}_0n$ for some constant $\tau_0$. As scattering enhances the rate of diffusion, we can expect the diffusion constant to be proportional to the scattering rate,
\be
D^{strob}(n) \sim n / \tau_0.
\ee
Note that this is in stark contrast to the conventional diffusion, where the diffusion constant  $D^{conv}$ is proportional to the scattering {\it time}, 
\be
D^{conv}(n) \sim \tau_0 /n,
\ee
see also Refs.~\cite{gravity, expansion}. Hence, we expect an inverse dependence of the diffusion constant on the scattering rate: scattering \textit{enhances} stroboscopic diffusion,
whereas it slows down diffusion conventionally. This heuristic is illustrated in Fig.~\ref{fig:BOsketch}.  Our goal in this section is an approximate derivation and asymptotic solution of the stroboscopic diffusion equation,
involving an explicit form of the diffusion constant.

\subsection{Derivation of stroboscopic diffusion}

In the following, we are aiming for an approximate analytic solution of the expansion problem.
Again, we assume that the force points in the diagonal direction of the lattice, ${\bf F} = F\,(1,1,...,1)^T$.
As in Ref.~\cite{gravity}, we assume the cloud to be translationally invariant in the perpendicular direction.
Equations~(\ref{eq:BOseteqns}) reduce to
\be
\dot{n} + \nabla j & = & 0 \label{eq:BOCoupledInhom} \\
\dot{e} + \nabla h - F\,j & = & 0 \n
\dot{j} +2 J^2 \nabla n + F \,e & = & - \tau_j^{-1} \,j \n
\dot{h} + \frac{3J^2}{d}\nabla e & = & -\tau_h^{-1} \, h.\nonumber
\ee
We assume two distinct scattering rates $1/\tau_j$ and $1/\tau_h$ for the damping of the particle and kinetic energy currents, respectively.
Note that these equations already break a complex integro-differential equation (the Boltzmann equation) down to four coupled partial differential equations in position space. 
Equations~(\ref{eq:BOCoupledInhom}) are still a rich and complicated set of coupled equations.
We cannot expect to be able to derive a single equation for particle density alone in a rigorous way, without neglecting parts of the physics.
We can nevertheless attempt to find an approximate equation capturing the dominant effects, involving several truncations. 
For the sake of clarity we give a summary of the following steps:
\begin{itemize}
\item[1.] We first study the system where the scattering rates $1/\tau_j$ and $1/\tau_h$ are constants. This makes the equations linear.
\item[2.] We allow for complex solutions. Due to linearity, the real and imaginary parts are separate solutions. 
\item[3.] We approximate $j$ in terms of $n$ and $e$, which results  in only three coupled equations.
\item[4.] We average over the fast time scale of Bloch oscillations. This decouples the dynamics of $n$ and $e,h$. We construct a real equation.
\item[5.] We finally substitute $\tau_j \rightarrow \tau_j(n)$, and $\tau_h \rightarrow \tau_h(n)$ in our simplified equations. 
These equations describe the stroboscopic dynamics  of $n(x,t)$. 
\end{itemize}

As we are interested in the strongly Bloch-oscillating regime, we expect the 
current mode $j$ to dominantly oscillate at  frequency $F$. Therefore, we substitute 
\be
j(x,t) & = & e^{i F t} \; \tilde{j}(x,t).
\ee
We assume that the dynamics of $\tilde{j}$ is much slower; it describes the spatial envelope of the oscillating current mode. 
After the substitution, the equation for $j$ in Eq.~(\ref{eq:BOCoupledInhom}) reads
\be
\dot{\tilde{j}} + i F \tilde{j} + 2 J^2 e^{-iFt} \nabla n + F \,e^{-iFt}  e & = & - \tau_j^{-1} \,\tilde{j}.
\ee
This equation contains drive and damping; we expect the long-time dynamics to be determined by $\dot{\tilde{j}}=0$,
which leads to the asymptotic solution
\be
j &\approx& \left(\frac{1}{iF + \tau_j^{-1}}\right)\left(-2J^2\nabla n + F e \right) 
\ee
Above, we have transformed $\tilde{j}$ back to $j$.
Note that the limit $F\tau \ll 1$ yields the current in the conventional hydrodynamic limit at high energies, see e.g. Ref.~\cite{gravity}.
However, as we are interested in the opposite limit of 
$F\tau \gg 1$, we use $1/(iF+\tau^{-1}_v)\approx -i/F - \tau_j^{-1}/F^2 $  to approximate 
\be
j &\approx& -i e + \frac{2 J^2 \tau_j^{-1}}{F^2} \nabla n - \frac{\tau_j^{-1}}{F}e + 2 i J^2 F^{-1} \nabla n 
\ee
The first term expresses the fact that without interactions, current and kinetic energies are related by the momentum shift of $\pi/2$; hence 
Bloch oscillations convert $j$ into $e$ periodically in time, and vice versa (see also Eq.~(\ref{eq:approx-bo}) below).
The second term expresses a real diffusive contribution to the current that is crucial. The third term describes the damping of the current mode
due to scattering, and the last term gives another imaginary contribution to the current that we discard in the following. 
As this truncation is done at the level of the current, the continuity equations still guarantee the conservation of energy and particle number.

Using our approximate result for the particle current, Eq~(\ref{eq:BOCoupledInhom}) simplifies to
\be
\dot{n} + \nabla \frac{2 J^2 \tau_j^{-1}}{F^2}\nabla n & = & i \nabla e \label{eq:three_eq} \\
\dot{e} + \nabla h - \tau_j^{-1}e + i F e & = & -\frac{2 J^2 \tau_j^{-1}}{F} \nabla n\n
\dot{h} + \frac{3J^2}{d}\nabla e & = & -\tau_h^{-1} \, h.\nonumber
\ee
In the absence of spatial gradients, the second equation describes the damping of Bloch oscillations in the kinetic energy,
\be
\dot{e}   - \tau_j^{-1}e + i F e \;  =  \;  0 \; +\;  {\cal O}(\nabla n, \nabla h), \label{eq:approx-bo}
\ee
which leads to a rapidly oscillatory motion of the kinetic energy. 
The kinetic energy therefore has the approximate solution $e(x,t)\sim e_0(x) e^{(iF -\tau^{-1})t}$. Therefore, the gradient of the kinetic energy
in Eq.~(\ref{eq:eq:three_eq}) averages out in time. This approximation is similar to the rotating wave approximation in quantum optics. 
Hence, particle diffusion approximately decouples from energy diffusion:
\be
\dot{n} + \nabla \frac{2 J^2 \tau_j^{-1}}{F^2}\nabla n & \approx &  0.
\ee
This is the approximate stroboscopic diffusion equation we were looking for. 
We  now study the corresponding system with a density-dependent diffusion constant, where $\tau(n)  = \tau_0/n$.
Defining  $D_0^{strob} = 2 J^2/(F^2 \tau_0)$, this equation can also be written as
\be
\dot{n} = D_0^{strob} \nabla \left(n\nabla n\right)\label{eq:nonDiff}.
\ee
It is important to realize that asymptotically, decoupling particle and energy diffusion in this limit does not violate any conservation
laws, as it would in Ref.~\cite{gravity}. Let us consider the long time limit, where all local kinetic energies have already been damped to $0$. 
The potential energy balance is still satisfied due to 
\be
\dot{E}_{\rm pot} & =  &   \int gx \; \dot{n}(x,t) \\
& = & \frac{1}{2}\int gx \; D_0^{strob}\nabla^2 n(x,t)^2 \n
& = &  \frac{1}{2}\int g(\nabla^2x) \; D_0^{strob} n(x,t)^2 = 0, \nonumber
\ee
where we have used partial integration. We  now present an asymptotic analytic solution of Eq.~(\ref{eq:nonDiff}).

\subsection{Scaling solution}
\label{sec:scaling}

\begin{figure}
 \begin{center}
  \includegraphics[width=\linewidth]{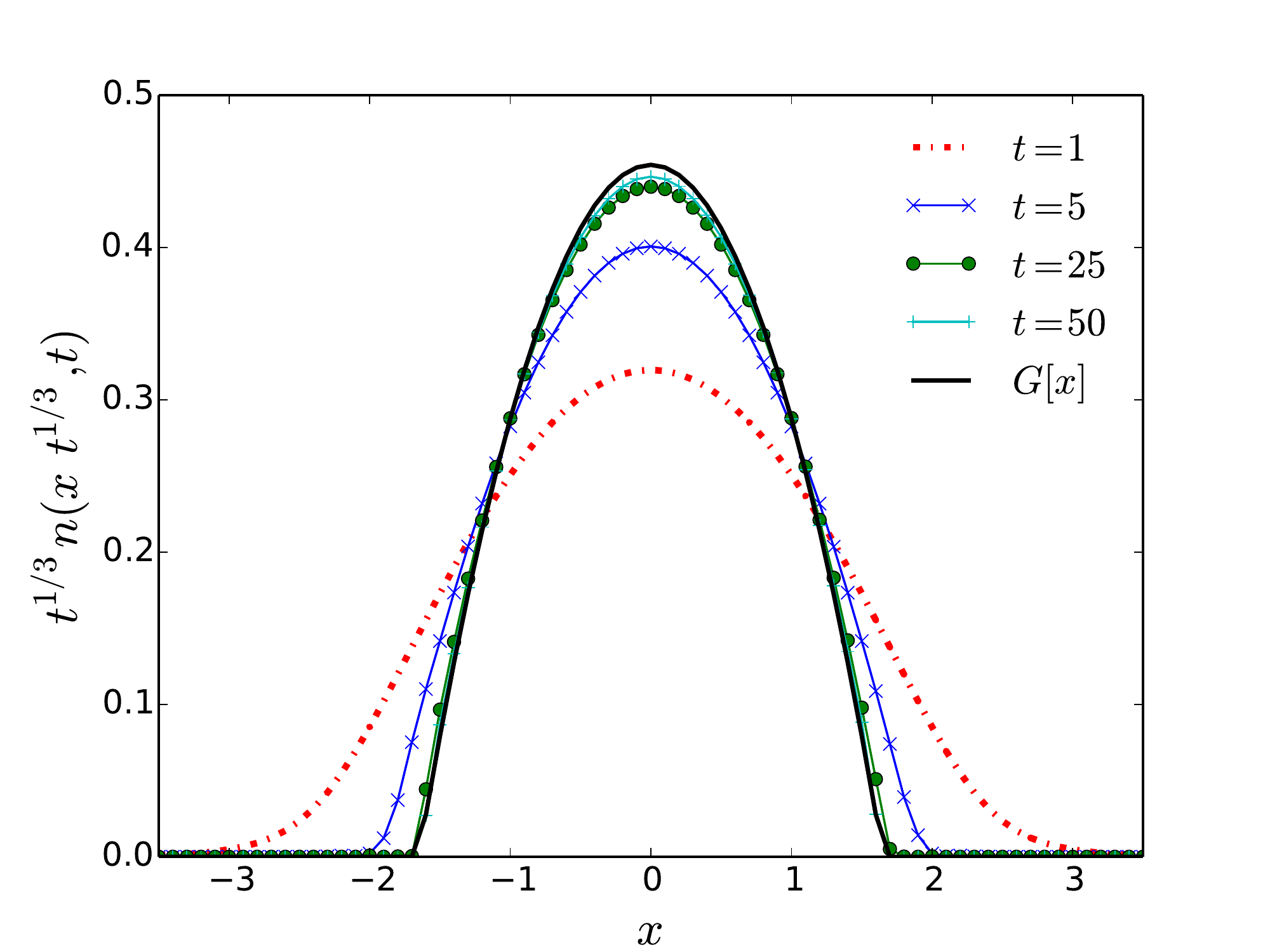}
  \caption{\CO Numerical solution of Eq.~(\ref{eq:nonDiff}) plotted against the scaling function $G[x]$. The simulated particle densities $n(x,t)$ were re-scaled 
as $t^{1/3}n(xt^{1/3},t)$ and plotted as a function of $x$ for different  times $t=1,5,25$ and $50$. The plot reveals that the numerical solutions $n$ 
assume the shape of the scaling function $G[x]$ upon re-scaling at long times.}
  \label{fig:scalingPlotNew}
 \end{center}

\end{figure}

As a final step, let us use a scaling ("Barenblatt") ansatz to obtain an asymptotic solution; see also~\cite{vazquez} and \cite{gravity}.  We use the following ansatz for the particle density:
\be
n(x,t) = \frac{1}{t^\alpha}G[x/t^\alpha]
\ee
We absorb $D_0^{strob}$ into the time variable, making Eq.~(\ref{eq:nonDiff}) dimensionless.
Combining the latter with the above scaling ansatz, and substituting $x$ by $z = x/t^\alpha$ yields
\be
0 & = & -t^{-1 - \alpha} \alpha G[z] -  t^{-1 - \alpha} z \alpha G'[z] \label{eq:scaling1}\\
 & & -  t^{-4 \alpha} G'[z]^2 -  t^{-4 \alpha} G[z] G''[z] \nonumber
\ee
The exponent of the time variable in the first term, coming from the time derivative, has to match the remaining terms with the slowest decay in time. 
It is actually possible to match all terms be setting  $\alpha = 1/3$.
This  result implies that the expansion is sub-diffusive. 

Let us now calculate the scaling function. Setting $\alpha = 1/3$ in Eq.~(\ref{eq:scaling1}) yields the ordinary differential equation
\be
G[z] + z G'[z] + 3 G'[z]^2 + 3 G[z]G''[z] = 0  \label{eq:scaling2}
\ee
A solution to this equation is given by
\be
G[z] = {\rm max} \{C_0 - z^2/6\,,\, 0 \} \label{eq:scalingform}
\ee
where $C_0$ is an arbitrary constant. 
In order to compare the scaling solution to our nonlinear diffusion Eq.~(\ref{eq:nonDiff}), we simulated the latter 
numerically, starting from a normalized Gaussian at time $0$. In Fig.~\ref{fig:scalingPlotNew} we compare the rescaled density profiles $t^{1/3} n(x\, t^{1/3},t)$ with the
scaling function $G[x]$. We chose $C_0 = 3^{1/3}/2^{5/3}\approx 0.4543$, which normalizes the integral of the scaling function to $1$. 
Fig.~\ref{fig:scalingPlotNew} shows that the rescaled densities assume the shape of
$G[z]$ at long times upon rescaling.  This  demonstrates the validity of the scaling law $x\sim t^{1/3}$ and the shape of the
scaling function.

Kolovsky\etal~\cite{kolovsky2010} also studied a Bloch-oscillating bosonic cloud and  approximated the many-body Schr{\"o}dinger equation by a lattice Gross-Pitaevskii equation
whereas we used the Boltzmann equation. Interestingly, a similar non-linear diffusion equation was derived whose diffusion constant scales as $D\sim n^2/F^2$ (we derived $D\sim n/F^2$ in this paper).
This led to a scaling law of $r\sim t^{1/4}$. Both approximations of the bosonic Schr{\"o}dinger equation have been used in the literature, they correspond to different energy domains
of the bosonic system, for more details we refer the reader to the seminal work by Gardiner\etal ~\cite{gardiner} and Zaremba\etal~\cite{zaremba}.

A subdiffusive scaling relation of $x\sim t^{1/3}$ was also found in Ref.~\cite{gravity}, which treated the opposite limit 
of $F\tau \ll 1$ (overdamped Bloch oscillations). Note, however, that the shape of the scaling function was different. The scaling analysis was carried out  for a coupled set of hydrodynamic equations for the particle end kinetic energy density. Here,
we heuristically derived the same scaling behavior in the regime of underdamped Bloch oscillations. An expanding cloud in an optical lattice with initially overdamped Bloch oscillations
will eventually enter the regime of underdamped Bloch oscillations as densities decrease, and hence scattering rates decrease over time. Our analysis suggests that both scaling
exponents are the same and that the scaling law will continue even when the cloud undergoes the transition between these two regimes.

\section{summary}
\label{sec:summary}

We have developed a new variational approach to solve the Boltzmann equation at high entropies for tight-binding systems. 
Our approach holds far beyond linear response; it is based on the linearization of the Boltzmann equation around a constant (maximum entropy) solution. 
It leads to a minimal set of coupled dynamic equations for momentum mode occupancies.

First, we have shown that in the limit of high scattering rates, the conventional hydrodynamic equations
can be recovered. 
Then we have presented two physical applications. (i) We studied the problem of the damping of Bloch oscillations in the Hubbard model,
which we approximated semiclassically in terms of a Boltzmann equation. 
We mapped this problem to the classical damped harmonic oscillator, providing analytic solutions
for the regimes of weakly damped, overdamped and marginally damped Bloch oscillations. 
For a quantitative comparison of our analytical results with the underlying theory, we have studied 
a one-dimensional discrete Boltzmann equation to explicitly calculate the relevant scattering rates and to allow for a full numerical simulation. 
(ii) We have then studied the problem of a strongly Bloch-oscillating, interacting cloud of fermions or incoherent bosons in a tilted lattice. 
While this problem was found to be too complex to solve without truncation, we presented an approximate solution in terms of a \textit{stroboscopic} diffusion equation
describing the dynamics of the cloud, time averaged over the fast Bloch oscillations. We have given a scaling solution for this simplified equation, leading to the
subdiffusive scaling relation $x\sim t^{1/3}$.

In the future, it will be intriguing to explore the class of dynamical problems that can be described in terms of the set of coupled equations that we derived.
As an example, our variational ansatz might be used to model the crossover from ballistic to diffusive dynamics of an expanding atomic cloud in an optical lattice,
which cannot be described in terms of a naive hydrodynamic approach alone. 

\section*{Acknowledgements}
I would like to thank Akos Rapp, Camille Aron, Achim Rosch, and David Huse for insightful discussions. This work was supported by the NSF MRSEC program through a Princeton Center for 
Complex Materials Fellowship (DMR-0819860).  I also acknowledge the support of the U.S. National Science Foundation I2CAM International Materials Institute Award, Grant DMR-0844115.

\newpage

\begin{appendix}
 \section{Derivation of Eq.~(\ref{eq:BOseteqns})}
 \label{sec:app1}
 We  now derive Eq.~(\ref{eq:BOseteqns}) line by line from Eq.~(\ref{eq:bracketCont}). We use  ansatz (\ref{eq:ansBOfk}) for $f$ in combination
 with the integral identities in Eq.~(\ref{eq:intIdentities}) and the orthogonality of the modes $1$, $\ek$, $\vk$ and $\ek\vk$ under the scalar product~(\ref{eq:BOdefScalar}).
This involves the fact that the momentum modes $\ek, \vk$ and $\ek\vk $ vanish under the momentum integral. We also use partial integration (P.I.) - note that there are no boundary terms due to
the periodicity of the Brillouin zone. This allows us to derive the following identities:
 \begin{widetext}
 \be
 \langle 1|\left(\partial_t + \vk \nabla_{\bf r} + {\bf F}\nabla_{\bf k}\right) \fk  \rangle & = & \int \frac{d \bk}{(2\pi)^d} (\partial_t + \vk \nabla_{\bf r} + {\bf F}\nabla_{\bf k})(n({\bf r},t) + \frac{e({\bf r},t)}{2J^2d}\ek + \frac{{\bf j}({\bf r},t)}{2 J^2}\vk + \frac{{\bf h}({\bf r},t)}{6 J^4}\ek\vk)\\
												& = & \dot{n} + \frac{1}{2 J^2}\int \frac{d \bk}{(2\pi)^d}  (\vk \nabla_{\bf r})({\bf j}({\bf r},t)\vk) \n
												& = & \dot{n}  +  \nabla_{\bf r}{\bf j}.\n
\langle \ek|\left(\partial_t + \vk \nabla_{\bf r} + {\bf F}\nabla_{\bf k}\right) \fk  \rangle & = & \int \frac{d \bk}{(2\pi)^d} \ek (\partial_t + \vk \nabla_{\bf r} + {\bf F}\nabla_{\bf k})(n({\bf r},t) + \frac{e({\bf r},t)}{2J^2d}\ek + \frac{{\bf j}({\bf r},t)}{2 J^2}\vk + \frac{{\bf h}({\bf r},t)}{6 J^4}\ek\vk)\\ 
												& = & \dot{e} \frac{1}{2 J^2 d}\int \frac{d \bk}{(2\pi)^d} \ek^2 + \frac{1}{6 J^4}\int \frac{d \bk}{(2\pi)^d} \ek (\vk\nabla_{\bf r})(\ek \vk {\bf h}) +  \frac{1}{2J^2} \int \frac{d \bk}{(2\pi)^d} \ek ({\bf F}\nabla_{\bf k}) (\vk {\bf j})\n												
												& \stackrel{P.I.}{=} & \dot{e}  +  \nabla_{\bf r}{\bf h} - {\bf F}{\bf j}.\n
\langle \vk|\left(\partial_t + \vk \nabla_{\bf r} + {\bf F}\nabla_{\bf k}\right) \fk  \rangle & = & \int \frac{d \bk}{(2\pi)^d} \vk (\partial_t + \vk \nabla_{\bf r} + {\bf F}\nabla_{\bf k})(n({\bf r},t) + \frac{e({\bf r},t)}{2J^2d}\ek + \frac{{\bf j}({\bf r},t)}{2 J^2}\vk + \frac{{\bf h}({\bf r},t)}{6 J^4}\ek\vk)\\ 
												& = & \frac{1}{2 J^2}\int \frac{d \bk}{(2\pi)^d} \vk (\vk \dot{{\bf j}}) + \int \frac{d \bk}{(2\pi)^d} \vk (\vk\nabla_{\bf r})n +  \frac{1}{2J^2d} \int \frac{d \bk}{(2\pi)^d} \vk ({\bf F}\nabla_{\bf k}) (\ek e)\n												
												& = & \dot{{\bf j}}  + 2 J^2 \nabla_{\bf r}n + {\bf F}e.\n
\langle \ek\vk|\left(\partial_t + \vk \nabla_{\bf r} + {\bf F}\nabla_{\bf k}\right) \fk  \rangle & = & \int \frac{d \bk}{(2\pi)^d} \ek\vk (\partial_t + \vk \nabla_{\bf r} + {\bf F}\nabla_{\bf k})(n({\bf r},t) + \frac{e({\bf r},t)}{2J^2d}\ek + \frac{{\bf j}({\bf r},t)}{2 J^2}\vk + \frac{{\bf h}({\bf r},t)}{6 J^4}\ek\vk)\n
												& = & \frac{1}{6 J^4}\int \frac{d \bk}{(2\pi)^d} \ek\vk (\ek\vk \dot{{\bf h}})  +  \frac{1}{2J^2d} \int \frac{d \bk}{(2\pi)^d} \ek\vk (\vk \nabla_{\bf r}) (\ek e)\\												
												& = & \dot{{\bf h}}  + \frac{3 J^2}{d}\nabla_{\bf r}e.\nonumber
 \ee
 \end{widetext}
 
\section{Scattering rates for the discrete one-dimensional Boltzmann equation}
\label{sec:AppCurrCurr}

We will now calculate the current-current matrix element of the linearized discrete one-dimensional Boltzmann equation~(\ref{eq:1DBoltz_1},\ref{eq:1DBoltz_2})
Trivially, scattering processes that simply exchange the incoming and outgoing momenta satisfy the energy and momentum constraint,
$ k_0,k_1 \rightarrow k_1,k_0 $. 
Obviously, this process does not lead to a damping of the particle current. However, as momentum is only defined modulo reciprocal lattice vectors,
Umklapp scattering processes are possible that relax the currents. 

\begin{figure}
 \begin{center}
  \includegraphics[width=.7\linewidth]{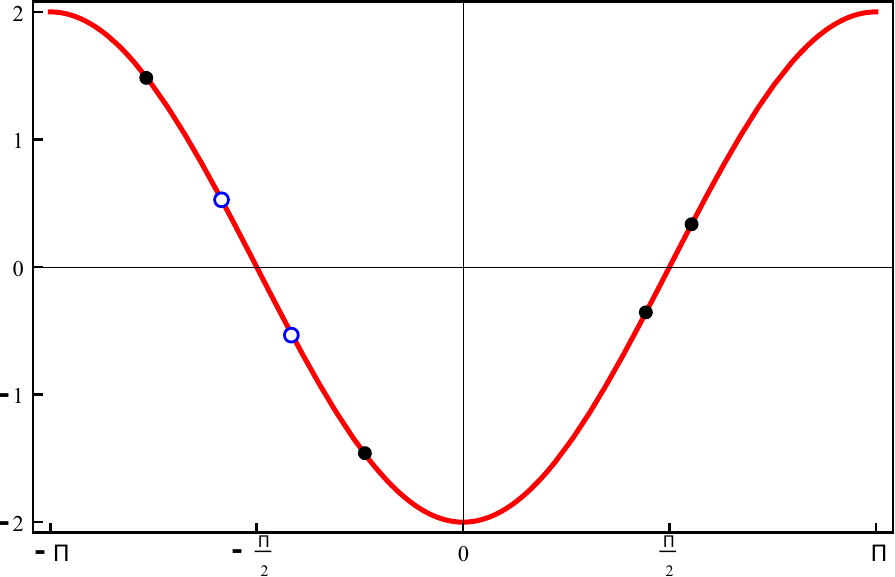}
  \caption{\CO Energy and momentum preserving scattering processes in one dimension. The red curve shows the kinetic energy as a function of momentum. Nontrivial scattering processes
are possible for pairs of momenta that are symmetrically centered around momentum $\pm \pi/2$ (blue circles). 
They can scatter in \textit{any} pair of final momentum states with 
 the same property (pairs of black points as two examples for a continuum of possibilities).}
  \label{fig:AppBOdispersion}
 \end{center}
\end{figure}

The discrete Boltzmann equation under consideration allows for several scattering processes that conserve the energy and quasi-momentum modulo $\pi$,  and relax the currents. 
These kinds of processes apply for very special ingoing and outgoing states; they are shown
in Fig.~\ref{fig:AppBOdispersion}. Consider two particles with momenta positioned symmetrically around the momentum $\pi/2$ or alternatively $-\pi/2$. These momentum states,
indicated by open blue circles in Fig.~\ref{fig:AppBOdispersion},  
have total kinetic energy $0$ and total quasimomentum $\pm \pi$:
\be
k_0 + k_1 & = & \pm \pi \label{eq:specialConstraint}\\
\epsilon_{k_0} + \epsilon_{k_1} & = & 0 \nonumber
\ee
This class of momentum states can now scatter in \textit{any} 
other pair of momentum states with the same properties (zero energy and total momentum $\pm \pi$). These pairs of momenta are
indicated as filled black circles in Fig.~\ref{fig:AppBOdispersion}.
It is important to realize that these are the only processes that can contribute to the relaxation of the current.
Note that these scattering states are a set of measure $0$ in the continuum theory (which is integrable), but they deliver a 
finite contribution to the scattering rates in our discrete model.

As the first step, we have to calculate the current-current matrix element of the linearized collision functional, given in Eq.~\ref{eq:matrixElements}.
As we are considering the discrete Boltzmann equation, we need to compute also the discrete version of this integral, which is given by
\be
\langle v_k|M|v_k\rangle  &=& C_0\sum_{k_0,k_1,k_2} \, (v_{k_0} + v_{k_1} - v_{k_2} - v_{k_0 + k_1 - k_2})^2\n
		&& \times \delta (\epsilon_{k_0} + \epsilon_{k_1} - \epsilon_{k_2} - \epsilon_{k_0 + k_1 - k_2})
\label{eq:AppBOvMvNewStart}
\ee
where $C_0 = n(1\pm n) U^2 /(4\,J\,N^2)$ and $\delta(\epsilon) = \delta_{\epsilon,0}$ is the discrete (Kronecker) delta. The different signs correspond to the case of bosons (+) and fermions (-). We already got rid of 
one summation by using the Kronecker delta for 
conservation of the quasi-momentum. Let us further simplify this sum. Equation~(\ref{eq:specialConstraint}) helps eliminate the energy constraint,
and it also implies that $v_{k_0 + k_1 - k_2}  =  v_{\pm \pi - k_2} = v_{k_2}$, which leads to 
\be
\langle v_k|M|v_k\rangle  & = &C_0\sum_{k_0,k_2} \, (2 v_{k_0} - 2 v_{k_2})^2 \label{eq:AppBOvMv1d}\\
			   & \stackrel{N \gg 1}{\longrightarrow} & n(1\pm n) \frac{U^2}{4\,J}\frac{1}{4\pi^2}\int dk_0\,dk_2\; (2 v_{k_0} - 2 v_{k_2})^2 \n
			    & = & 4\,n(1\pm n) U^2\,J \nonumber
\ee
Above, we have approximated the discrete sum in the absence of the delta-constraint by a continuous integral. This approximation works very well, as we have also calculated
the discrete matrix element $\langle v_k|M|v_k\rangle$ numerically using Eq. (\ref{eq:AppBOvMvNewStart}), and found excellent agreement with (\ref{eq:AppBOvMv1d}) already
for $N=20$.

\end{appendix}

\end{document}